\documentclass[reprint,aps,prc,notitlepage,amsmath,dvipsnames,nofootinbib,superscriptaddress,doi=true,preprintnumbers]{revtex4-1}

\usepackage[utf8]{inputenc}
\usepackage[T1]{fontenc}
\usepackage{lmodern}
\usepackage{microtype}
\usepackage[english]{babel}
\usepackage[hyperindex=true,colorlinks=true]{hyperref}

\usepackage{graphicx}
\usepackage{rotating}
\usepackage{pifont}
\usepackage{xcolor}

\usepackage{amsmath}
\usepackage{amssymb}
\usepackage{amsfonts}
\usepackage{bm}
\usepackage{xfrac}
\usepackage{braket}

\usepackage{url}

\usepackage{lmodern}
\usepackage{microtype}

\newcommand{\ra}{\rangle}

\newcommand{\opt}{NNLO$_\mathrm{opt}$ }
\newcommand{\NNLOsim}{NNLO$_\mathrm{sim}$ }

\def\orcid#1{\kern .08em\href{https://orcid.org/#1}{\includegraphics[keepaspectratio,width=0.7em]{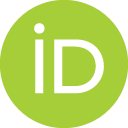}}}

\begin{document}

\title{Quantum Monte Carlo in Configuration Space with Three-Nucleon Forces}

\author{Pierre Arthuis \!\!\orcid{0000-0002-7073-9340}}
\email{parthuis@theorie.ikp.physik.tu-darmstadt.de}
\affiliation{Technische Universit\"at Darmstadt, Department of Physics, 64289 Darmstadt, Germany}
\affiliation{ExtreMe Matter Institute EMMI and Helmholtz Forschungsakademie Hessen für FAIR (HFHF), GSI Helmholtzzentrum für Schwerionenforschung GmbH, 64291 Darmstadt, Germany}
\author{Carlo Barbieri \!\!\orcid{0000-0001-8658-6927} }
\email[]{carlo.barbieri@unimi.it}
\affiliation{Dipartimento di Fisica, Università degli Studi di Milano, Via Celoria 16, I-20133 Milano, Italy}
\affiliation{INFN, Sezione di Milano, Via Celoria 16, I-20133 Milano, Italy}
\author{Francesco Pederiva } 
\affiliation{Physics Department, University of Trento, Via Sommarive 14,  I-38123 Trento, Italy}
\affiliation{INFN-TIFPA Trento Institute of Fundamental Physics and Applications, Via Sommarive, 14, 38123 Povo TN, Italy}
\author{Alessandro Roggero \!\!\orcid{0000-0002-8334-1120}\, }
\affiliation{Physics Department, University of Trento, Via Sommarive 14,  I-38123 Trento, Italy}
\affiliation{INFN-TIFPA Trento Institute of Fundamental Physics and Applications, Via Sommarive, 14, 38123 Povo TN, Italy}
\affiliation{InQubator for Quantum Simulation (IQuS), Department of Physics,University of Washington, Seattle, WA 98195, USA}

\date{\today}

\begin{abstract}
Neutron matter, through its connection to neutron stars as well as systems like cold atom gases, is one of the most interesting yet computationally accessible systems in nuclear physics. The Configuration-Interaction Monte Carlo (CIMC) method is a stochastic many-body technique allowing to tackle strongly coupled systems. In contrast to other Quantum Monte Carlo methods employed in nuclear physics, the CIMC method can be formulated directly in momentum space allowing for an efficient use of non-local interactions.  
In this work we extend CIMC method to include three-nucleon interactions through the normal-ordered two-body approximation. We present results for the equation of state of neutron matter in line with other many-body calculations that employ low resolution chiral interactions, and provide predictions for the momentum distribution and the static structure factor.
\end{abstract}

\preprint{IQuS@UW-21-026}

\maketitle

\section{Introduction}

A common feature among effective theories of complex systems is their intrinsic non-locality originating from the integration of non-essential degrees of freedom. This feature permeates nuclear physics at different levels because, besides the many-body mechanisms at play, the relevant components of the nucleus (protons and neutrons) are themselves the result of the underlying degrees of freedom of Quantum Chromodynamics (QCD). For this reason, modern ‘first principle’ theories of nuclear systems rely on effective field theories (EFTs) to construct realistic inter-nucleon interactions~\cite{Hamme2020RevMdPhys}. In this approach, the exchange of light mesons (pions) and the inclusion of contact counterterms are ordered according to a perturbative series~\cite{Piarulli2020Front, Epelbaum2020Front, Entem2020Front}. However, most many-body methods used to study low energy processes cannot deal explicitly with mesonic fields. Mesons are then integrated out using some regulator that imposes a cutoff in momentum space, and the resulting potential is in general correctly described only including explicit momentum-dependent terms.

Partly due to the recent success of the application of Quantum Monte Carlo in dealing efficiently with Hamiltonians including complex operatorial dependencies, like the Auxiliary Field Diffusion Monte Carlo (AFDMC)~\cite{Carlson2015rmp,Lonardoni2018}, a substantial effort was made in order to produce local versions of chiral effective interactions (local $\chi$-EFT potentials)~\cite{Gezerlis2014,Lynn2019}. These potentials, while retaining some of the spirit of the chiral EFT, are plagued by a number of necessary inconsistencies (such as in the introduction of regulators that break Fierz symmetry), making the basic connection with the QCD symmetries feebler and feebler. 

In this context the use of methods that can smoothly work in momentum space, avoiding the necessity of this further step of making the interaction local, would be preferable. 
Non-local Hamiltonians can be easily handled for finite nuclei by exploiting methods formulated in a Fock space (i.e., the space of the Slater determinants and including different particle numbers) built on sets of localised basis functions~\cite{Hergert2020}.  Several \emph{ab initio} methods, such as Many-Body Perturbation Theory (MBPT)~\cite{Tichai2020a}, Self-Consistent Green's Function (SCGF)~\cite{Barbieri:2016uib,Soma2020a}, Coupled Cluster (CC)~\cite{Hagen2015,Hagen2016a} or the In-Medium Similarity Renormalization Group (IMSRG)~\cite{Hergert2016,Stroberg:2016ung}, can indeed provide results beyond light nuclei and have reached masses above A$\sim$100~\cite{Binder:2013xaa,Morris2018,Arthuis2020,Miyagi2021} up to first estimations of $^{208}$Pb~\cite{Hu2021}. Working in momentum space also gives direct access to quantities such as the momentum distribution or the static structure factor that are not so easily computable in coordinate space, and that are an important ingredient for the estimate of further observables of interest. Moreover, momentum space calculations facilitates accurate determinations of optical potentials~\cite{Idini2019}, that would draw a bridge between \emph{ab initio} methods and the description of dynamical processes for medium-heavy nuclei.

For the case of infinite matter, the SCGF approach can be implemented directly in momentum space, which allows to handle high-momentum components and therefore works equally well with both soft and hard interactions~\cite{Soma2008Tmtx3NF,Carbone2013TNF}. Moreover, the moderate computing requirements (sometimes not even requiring parallelisation) and its general formulation at finite temperatures make SCGF the method of choice to investigate a large range of phenomena, such as nucleon propagation in the medium~\cite{Rios2021MnFrPath} or pairing effects~\cite{Ding2016GapsNM} and temperature dependence~\cite{Carbone2018LiqGas,Carbone2020pnmT} of neutron matter. In spite of their great versatility, current implementations of SCGF face some limitations in the pairing instabilities for symmetric matter at very low temperatures and densities, and in the precision of response functions that would require effective vertices to go beyond resummations of dressed ring diagrams. While some step could be taken using a Nambu covariant formalism~\cite{Drissi2021NmbCovSCGF}, these problems may be solved more efficiently using a direct digonalization in the full Fock space that goes beyond common post-Hartree-Fock many-body truncations.

The ground state properties of neutron matter, from saturation to very low densities, are particularly interesting due to its vicinity to the unitary Fermi gas limit~\cite{Vidana2021Front}. In this regime, neutron matter shares very similar properties with cold gasses at the Feshbach resonance~\cite{Chin2010rmp}. The resulting equation of state (EoS) directly affects the structure of the inner core of neutron stars~\cite{Chamel2008NS} and the skin of heavy neutron-rich isotopes~\cite{Vinas2014}. Hence, high-accuracy investigations with the best advanced nuclear interactions are of remarkable importance.

A few years ago, Roggero et al.~introduced a flavor of Quantum Monte Carlo simulations, under the name of Configuration-Interaction Monte Carlo (CIMC)~\cite{Mukherjee2013,Roggero2013,Roggero2014,Roggero2015}, partly originating from existing Shell Model Monte Carlo algorithms~\cite{Koonin1997}, providing an efficient way to expand an arbitrary state in the Fock space and stochastically propagate it in imaginary time (IT). Carrying out the propagation for a sufficiently long IT gives the possibility of sampling the expansion of the ground state. This method combines the natural language needed to deal with momentum-dependent interactions to the efficiency of Quantum Monte Carlo techniques.  The method uses CC amplitudes as a trial wave function to guide the IT propagation and therefore it is computationally more expensive than CC and SCGF computations~\cite{Hagen2014,McIlroy2019} embedded in the same Fock space. However, it is not limited by the same many-body truncations of these schemes and it is guaranteed to improve toward the correct ground state while satisfying the variational ansatz.  The method demonstrated to be very efficient for Hamiltonians limited to two-body interaction. However, the extension to three-body forces (absolutely necessary for a realistic description of nuclear systems of interest) was hindered by technical limitations.

In this paper we present the first CIMC results obtained for cold, catalyzed neutron matter interacting through a $\chi$-EFT potential that includes full three-nucleon forces (3NFs).
We exploit the \opt~\cite{Ekstrom:2013kea} interaction for the two-nucleon sector and apply our method at different densities below and above the nuclear saturation density $\rho_0=0.16$.
The propagation is started from the amplitudes obtained in the second-order Møller-Plesset (MP2) solution for the same Hamiltonian. Besides the energies and the related EoS of neutron matter, we illustrate results for the momentum distribution and the static structure factor, compared to calculations with other methods.

The paper is organized as follows. In Sec.~\ref{sec:method} the CIMC method is reviewed and extended to the case of 3NFs. Some technical details of how we store nuclear matrix elements are pivotal to practical implementations. We collect them in App.~\ref{sec:ME_storage}.
Secs.~\ref{sec:EoS},~\ref{sec:k_dist} and~\ref{sec:Sk} present the results for the EoS, the momentum distribution and the static structure factor, respectively. Sec.~\ref{sec:conclusions} is devoted to conclusions.

\section{Method}
\label{sec:method}

Configuration-interaction (CI) approaches have been widely used over the past decades in atomic, molecular and nuclear physics, describing a variety of systems and observables. They rely on the expansion of the $N$-body wave function $|\Psi^\text{FCI}\ra$ in the space spanned by $N$-particle Slater determinants $|\Phi_i\ra$,
\begin{equation}
    |\Psi^\text{FCI}\ra \equiv \sum_i C_i |\Phi_i\ra \ ,
\end{equation}
where $C_i$ corresponds to the expansion coefficient associated to the determinant $|\Phi_i\ra$. Those coefficients are ultimately obtained by diagonalising the Hamiltonian of the system in the truncated model space. The combinatorial cost in the size of the model space and the number of particles 
can make such an approach unusable for systems larger than a few-particles. For nuclear physics, standard CI and Quantum Monte Carlo reach such a wall for around \hbox{$A\approx$ 12--20} nucleons. One way to reduce this cost has been
%
%
the development of novel Monte Carlo methods, like Auxiliary-Field Monte Carlo in configuration space~\cite{Lang1993,Koonin1997,Alhassid2001}, or more recently in quantum chemistry the Full CI Quantum Monte Carlo approach~\cite{Booth2009,Cleland2010,Booth2010,Overy2014,Neufeld2018,Blunt2019}.

Configuration-Interaction Monte Carlo~\cite{Mukherjee2013} is another of these approaches, already successfully applied to the homogeneous electron gas~\cite{Roggero2013}, small molecules~\cite{ROGGERO2018241} and neutron matter~\cite{Roggero2014,Roggero2015}. One starts by defining a suitable projector $\mathcal{P} = \exp\left(-\tau (H - E_\text{T})\right)$, where $H$ is the Hamiltonian of the system, $\tau$ a finite step in imaginary time and $E_\text{T}$ a shift in energy. The ground-state wave-function $|\Psi\ra$ is then extracted by applying $\mathcal{P}$ iteratively $N_\tau$ times on an initial state $|\Phi\ra$,
\begin{equation}
    |\Psi\ra \equiv \lim_{N_\tau \to \infty}  \mathcal{P}^{N_\tau} |\Phi\ra \ ,
\end{equation}
where $|\Phi\ra$ has been selected to have a large overlap with the ground state. 

The evolution is then done stochastically, with probabilities proportional to the matrix elements of the propagator $\mathcal{P}$ between different Slater determinants. As such, it depends on the interaction part of the Hamiltonian and not just on its kinetic part like for coordinate-space Monte Carlo methods. Since this method is not exempt from the sign problem one introduces a guiding wave function $|\Phi_G\ra$, ideally pre-determined from a low-cost method. On configurations for which the wave function does not cancel out, one can then introduce an auxiliary family of Hamiltonians $H_\gamma$ with $\gamma \in [0;1]$, designed in such a way to keep the wave-function positive semi-definite~\cite{Mukherjee2013}. This allows for an evolution free of the sign problem, and ultimately CIMC provides an upper bound on the ground-state energy of the system.

The calculations are done in the space of Slater determinants spanned from a set of discretized single-particle eigenstates of momentum and of spin and isospin projections.
These basis states form a cubic lattice that corresponds to a box with periodic boundary conditions in coordinate space. The  density of the system $\rho$ and the number of particles included determine the side length of the box and the lattice spacing in momentum space. 
A cutoff is imposed on the maximal squared momentum allowed $k^2_{\mathrm{max}}$, and convergence is reached once the difference in energy when increasing the cutoff is smaller than the statistical uncertainty. This proved to be the case for all present calculations at $k^2_{\mathrm{max}} = 24$ in squared units of the momentum space lattice. Though the construction of the Hamiltonians $H_\gamma$ would require to extrapolate to $\gamma = -1$ to get to the exact value, we noticed during exploratory calculations that for densities around and below nuclear saturation, the differences when varying $\gamma$ were smaller than the statistical uncertainties, and results can be considered converged. For larger densities, linear extrapolations from results obtained with values of $\gamma$ comprised between 0 and 1 yield at most a correction of 50~keV/N for $\rho = 0.32\ \mathrm{fm}^{-3}$, in line with previous results obtained with two-body potentials for neutron matter and symmetric nuclear matter~\cite{Rrapaj2016}. This indicates that the sign problem is very mild in our system, and we thus discuss only results with $\gamma = 0$ in the following.

One of the key aspects for CIMC calculations is the choice of the guiding wave function $|\Phi_G\ra$. The closer to the true ground state it is, the better the final results, but this comes at the price of a costlier pre-processing. For the present calculations, $|\Phi_G\ra$ was obtained by computing the MP2 energy and amplitudes. This is equivalent to applying the coupled cluster with doubles (CCD) method and stopping after the first iteration, and provides a good enough approximation for a perturbative system like neutron matter.

While previous implementations of CIMC were restricted to two-body (NN) forces, we presently extend the method to include effects from three-body forces, which have been shown to be critical to the reproduction of nuclear matter saturation~\cite{Hebeler2011,Hebeler2021} as well as for reproduction of e.g.~driplines in finite nuclei~\cite{Otsuka2010,Cipollone2013}. A common strategy adopted within the nuclear physics community is to normal-order $H$ with respect to the reference state and then truncate at the two-body level, resulting in the NO2B approximation~\cite{Hagen2007,hebeler09a,Roth:2011vt,HoltKaiser2017,Frosini2021}.
Focusing on the three-body part of the original Hamiltonian $H$,
\begin{equation}
    H_3 = \frac{1}{36} \sum_{pqrstu} \bar{v}_{pqrstu}
    a^\dagger_p a^\dagger_q a^\dagger_r a_u a_t a_s \ ,
\end{equation}
where $\bar{v}_{pqrstu}$ are the antisymmetrized three-body matrix elements of $H$, and $\{a^\dagger_p, a_p\}$ are particle creation and annihilation operators, this consists in rewriting it in normal order with respect to the HF reference state, reading as
\begin{equation}
\label{eq:normal_order}
\begin{split}
    H_3 &= \frac{1}{6} \sum_{ijk} \bar{v}_{ijkijk}
+ \frac{1}{2} \sum_{ijpq} \bar{v}_{ijpijq} :a^\dagger_p a_q: \\
&\phantom{=} + \frac{1}{4} \sum_{ipqrs} \bar{v}_{ipqirs} 
:a^\dagger_p a^\dagger_q a_s a_r: \\
&\phantom{=} + \frac{1}{36} \sum_{pqrstu} \bar{v}_{pqrstu} :a^\dagger_p a^\dagger_q a^\dagger_r a_u a_t a_s: \,
\end{split}
\end{equation}
where :AB: represents the operator AB in normal order and $i,j,k$ are occupied states. One then discards the last term in Eq.~\eqref{eq:normal_order}, known as residual three-body forces.
This has proved to yield a very good reproduction of systems at the price of a small truncation error, especially for neutron matter~\cite{Hagen2014,Drischler:2017wtt}. As such, the present calculations rely on the use of the NO2B Hamiltonian, with implementation details being discussed in App.~\ref{sec:ME_storage}.

\section{Results}

\begin{figure}[t]
  \includegraphics[width=\linewidth]{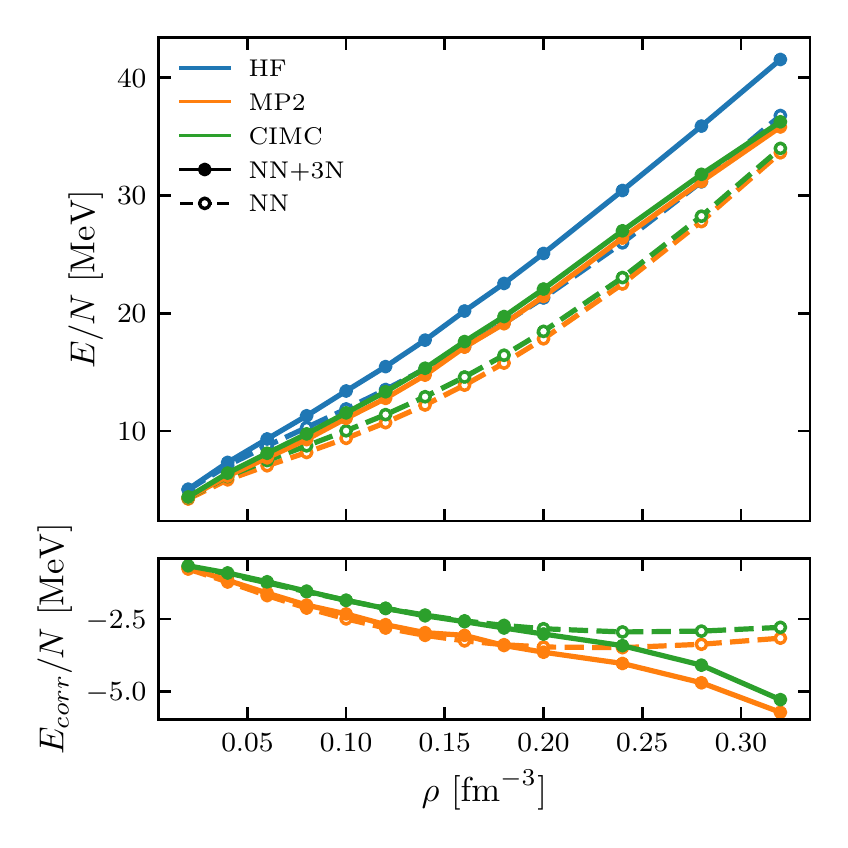}%
  \caption{\label{fig:EoS_NO2B} 
  \emph{Top panel.} Equation of state of neutron matter computed from CIMC in a periodic box with 66 neutrons and compared to results from HF mean field and second-order perturbation theory.
 Dashed and full lines show predictions for the sole two-nucleon \opt interaction and with the addition of 3NFs at NNLO, respectively.
 Note that the NN-only HF result (dashed blue line) is partially hidden as it overlaps with the MP2 and CIMC curves for full 3NFs.
  \emph{Bottom panel.} Correlations energies per particle.
  }
\end{figure}

We compute pure neutron matter using the CIMC framework discussed in the previous section and the \opt Hamiltonian introduced in Ref.~\cite{Ekstrom:2013kea}. This interaction is optimized with respect to data in the two-nucleon sector but 3NFs are still required to reproduce the EoS for nucleonic matter and the driplines in neutron-rich isotopes. Following Ref.~\cite{Ekstrom:2013kea} we add a next-to-next-to-leading-order (NNLO) three-nucleon interaction with values of $c_D = -0.20$ and $c_E = -0.36$ for the low-energy coupling constants, and apply local regulators depending on momentum transfer~\cite{Navratil2007c} as discussed in~\cite{Hagen2014} with a value of $\Lambda$~=~500~MeV/c.

For all results shown below we consider a periodic box containing 66 neutrons, which is known to be the optimal choice to minimize the finite size effects while still requiring moderate computational costs. We have also checked that our results are stable with respect to the number of particles included and computations resulted to be largely converged with respect to $k^2_{\mathrm{max}}$, as discussed above.

In the following, we demonstrate results for different properties of interests for neutron matter---namely the EoS, the momentum distribution, and the static response---and discuss the effects arising form the inclusion of 3NFs.

\subsection{Equation of State}
\label{sec:EoS}

The energy per particle as a function of density is displayed in Fig.~\ref{fig:EoS_NO2B}. This shows the three steps of the CIMC calculations: the energy from the Hartree-Fock approximation, the second-order Møller-Plesset perturbation theory, and finally the prediction from Monte Carlo diffusion. As expected for neutron matter, the bulk of the correlation is already captured by a perturbative expansion but MP2 still over binds slightly. The CIMC algorithm corrects this behaviour and at the same time provides a solid variational upper bound to the ground-state energy. It will be interesting to revisit this with harder Hamiltonians or especially in symmetric nuclear matter, where CIMC typically corresponds to resumming high-rank particle-hole excitations~\cite{Rrapaj2016}.
The 3NFs have the overall effect of making the system less bound across the whole range of densities considered. Nevertheless, the bulk of this repulsion originates solely in the mean-field step. This is demonstrated by the correlation energies $E_{corr} \equiv E - E_{HF}$ reported in the bottom panel of Fig.~\ref{fig:EoS_NO2B}. Many-body correlations are not appreciably affected by 3NFs for dilute systems up to $\sim 1.2\rho_0$ and instead generate more attraction at large neutron densities (although not sufficiently strong to invert the repulsion generated by 3NFs themselves at the mean-field level). The moderate contribution of the 3NF can be traced back to the design of the \opt Hamiltonian, made to minimise contributions beyond the two-nucleon force~\cite{Ekstrom:2013kea} and the use of a local cutoff~\cite{Holt2020}.

\begin{figure}[t]
  \includegraphics[width=\linewidth]{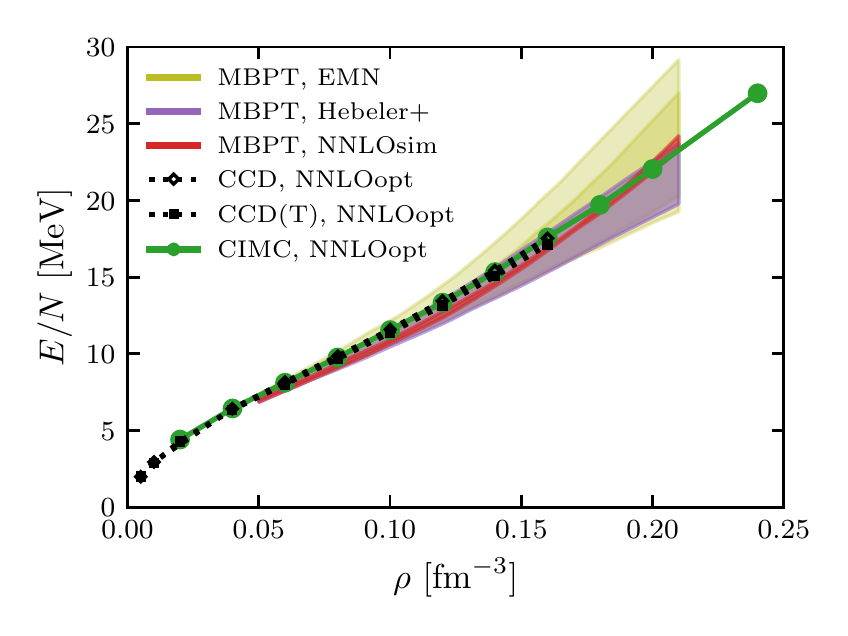}%
  \caption{\label{fig:EoS_chiral} Equation of state of neutron matter for the NO2B \opt Hamiltonian with 66 neutrons compared with CCD with NO2B and CCD(T) with full 3NF results with \opt~\cite{Hagen2014} and MBPT calculations at fourth order with various chiral forces: EMN at NNLO and N3LO, \NNLOsim at NNLO and Hebeler+ at N3LO (see~\cite{Drischler:2017wtt} and references therein). For EMN, the band corresponds to the calculated theoretical uncertainty, the two shades corresponding to NNLO and N3LO. The bands for Hebeler+ and \NNLOsim correspond to a variation of cutoffs and/or renormalization scale.}
\end{figure}

Fig.~\ref{fig:EoS_chiral} benchmarks our results including 3NFs against other computations available in the literature. The CCD and CCD(T) are obtained from the same (two-nucleon) \opt Hamiltonian but with slightly different choices for the low-energy constants $c_D$ and $c_E$ in the 3NF sector~\cite{Hagen2014}. Colored bands are uncertainty estimates from fourth-order MBPT calculations for different chiral forces~\cite{Drischler:2017wtt}.
The present CIMC results fit well within the prediction of all chiral Hamiltonians shown, which gives further confidence in the predictive capabilities of this method.  Particularly remarkable is the fact that CIMC points are effectively on top of the CCD results and CCD(T) curves that are obtained with a nearly identical interaction.

\subsection{Momentum distribution}
\label{sec:k_dist}

\begin{figure}
  \includegraphics[width=\linewidth]{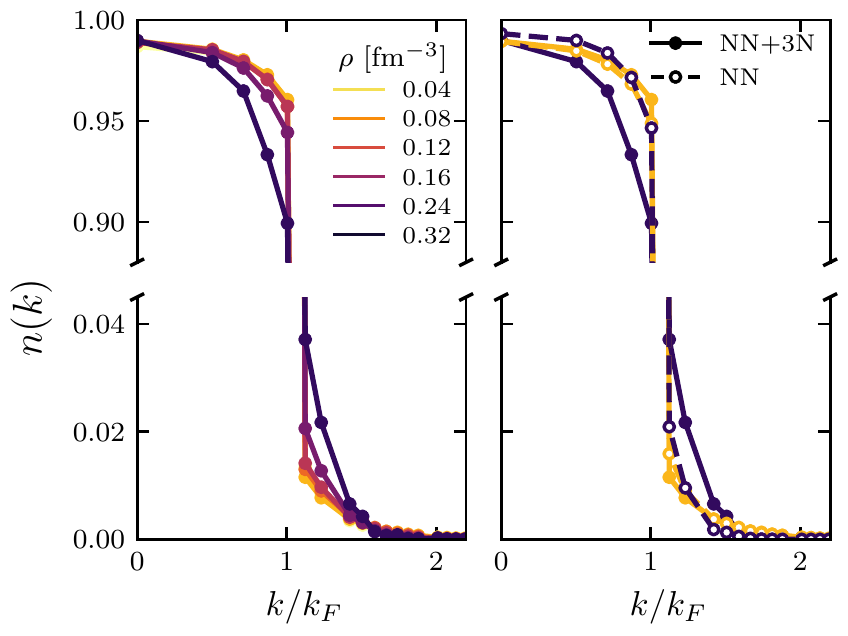}%
   \caption{\label{fig:mom_dist}
   \emph{Left panel}. Momentum distribution of neutron matter for the \opt Hamiltonian plus 3NFs, computed in a periodic box with 66 particles, for different densities. 
\emph{Right panel}. Comparison of $n(k)$ computed with and without 3NFs for two cases, at half and twice saturation density.
The momentum is expressed as a fraction of the Fermi momentum~$k_F$.
}
\end{figure}

Having gained confidence in our CIMC approach with 3NFs, we consider results for the momentum distribution $n(k)$~\cite{Furnstahl2002}. 
For fermionic systems, deviations from the ideal Fermi-Dirac distribution inform on beyond-mean-field correlations induced by a particular Hamiltonian. The occupation of states as a function of their momentum $k$ is obtained directly from the Quantum Monte Carlo walkers for each lattice point in the single-particle model space.

The momentum distribution of neutron matter is displayed in the left panel of Fig.~\ref{fig:mom_dist} for various densities and the full Hamiltonian that includes 3NFs.
The depletion at zero momentum is found to be between 1-2\% and it is independent of the density. Such small effect is in part due to the fact that neutron matter is generally less correlated than symmetric nuclear matter~\cite{Rios2020Front}. At the same time, we observe a weaker effect compared to predictions from other standard chiral N3LO forces in the literature~\cite{Rios2014highk,Carbone2013TNF,Rios2020Front}, showing that the \opt interaction is particularly soft. 

The left panel of Fig.~\ref{fig:mom_dist} shows that the $n(k)$ curve is almost independent from $\rho$ up to saturation density. On the contrary, correlation effects increase for $\rho\approx2\rho_0$ and smooth the discontinuity in occupations at the Fermi surface. The origin of this behavior can be traced to 3NFs and be better understood from the right panel, were we compare the momentum distributions computed with and without 3NFs. At low density, the bulk of correlations comes only from the two-body \opt Hamiltonian and the two curves are on top of each other. For densities $\rho> $0.20~fm$^{-3}$ the two-nucleon Hamiltonian provides a distribution very close to the one observed a lower densities but the contribution of 3NFs becomes dominant.

Computations based on SCGF have pointed out that, around saturation density, 3NFs have a quantitatively small impact on $n(k)$ and the fragmentation properties of single-particle strength, even though they have a large influence on the energetics and thermodynamics of the system~\cite{Carbone2013TNF}.
Our results from Figs.~\ref{fig:EoS_NO2B} and~\ref{fig:mom_dist} confirm this picture but suggest that correlations from 3NFs become relevant at large densities.

\subsection{Static Structure factor S(q)}
\label{sec:Sk}

The static structure function $S(q)$ carries information about the response of the system to density excitations with  momentum transfer equal to $|\mathbf{q}|=q$. It can be thought as the energy average of the (vector) dynamic structure factor $S(q,\omega)$ defined as
\begin{equation}
S(q,\omega) = \sum_n \left|\langle \Psi_n\lvert\rho(q)\rvert\Psi_0\rangle\right|^2\delta(E_n-E_0-\omega)\;,
\end{equation}
with $\rvert\Psi_n\rangle$ the energy eigenstates of the nuclear Hamiltonian with eigenvalue $E_n$ and $\rho(q)$ the Fourier transform of the density operator. This quantity is proportional to the differential scattering cross section of processes coupling to the density of the system and transfering energy $\omega$ and momentum $q$. Upon integration over the energies, the static structure factor can be then expressed as
\begin{equation}
S(q) = \int_0^\infty d\omega S(q,\omega) = \langle\Psi_0\lvert\rho(q)^\dagger\rho(q)\rvert\Psi_0\rangle\;.
\end{equation}
These quantities, together with the corresponding spin (or axial) responses, carry important information about neutrino scattering from neutrons in infinite matter. Previous calculations of $S(q)$ where performed in the high-temperature small-density regime employing either a virial expansion~\cite{HOROWITZ2006326} or lattice Monte Carlo methods exploiting the similarity between low-density neutron matter and a unitary Fermi gas which present no sign problem~\cite{PhysRevC.101.045805}. Using the CIMC we were able to extend these earlier calculations and estimate the density static structure factor $S(q)$ at zero temperature and large densities, beyond the reach of either method. This is an important step forward to characterize the response of neutron matter to neutrinos using \emph{ab initio} methods with realistic NN and 3N interactions from chiral EFT. 

\begin{figure}[t]
  \includegraphics[width=\linewidth]{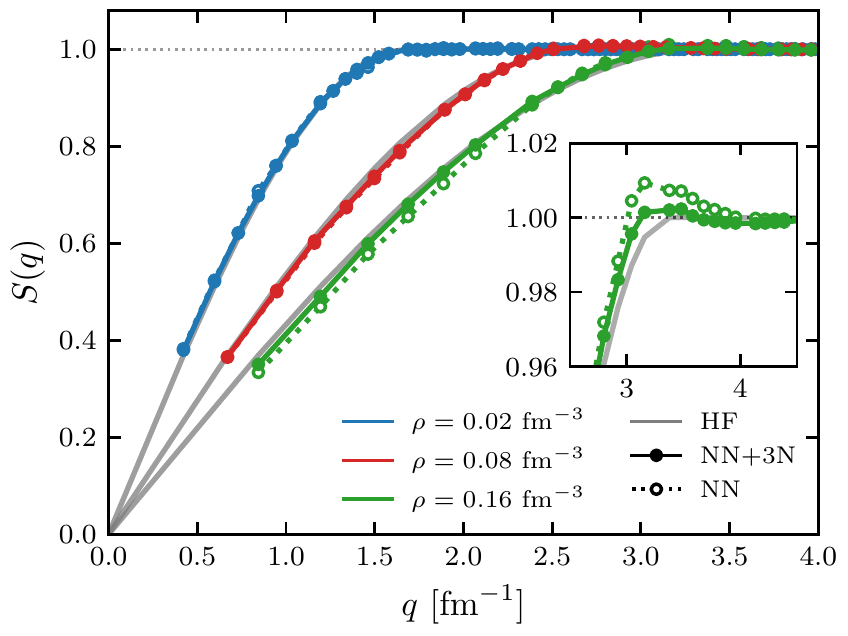}
  \caption{\label{fig:sofq} Structure function of neutron matter for the \opt Hamiltonian with 66 neutrons for various densities with and without 3NF. The inset shows more detail around $q=2k_F$ for saturation density $\rho=0.16\;\rm{fm^{-3}}$. We employ a correction for finite size effects as described in Appendix~\ref{sec:Sk_formalism}.}
\end{figure}

We present our results for $S(q)$ in Fig.~\ref{fig:sofq} for densities ranging from $\rho=0.02\;\rm{fm^{-3}}$ (shown in blue) to nuclear saturation density (shown in green). A detailed description of our procedure to estimate $S(q)$ from the configurations sampled by CIMC is provided in Appendix~\ref{sec:Sk_formalism}. The solid grey lines are results obtained within the Hartree-Fock approximation which in this case is equivalent to the free Fermi gas result. We find a rather weak effect from interactions of the order of $\approx 1\%$ at nuclear saturation density. These results are compatible with the response to density excitations being dominated by the repulsive part of the interactions and it does not show a pronounced peak, which would instead be expected for purely attractive forces. As one can see in more detail in the inset, showing results for $\rho=0.16\;\rm{fm^{-3}}$ only, the net effect of adding 3NFs is to further suppress the oscillations in $S(q)$ and is consistent with the net repulsion brought by these interactions. 

In future work it would be interesting to also explore the spin (axial) response of neutron matter as this will provide information about neutrino properties in bulk neutron matter like their mean free path and emissivity~\cite{Shen2013,Riz_2020}.

\section{Conclusions}

In this work we have extended the Configuration-Interaction Monte Carlo method for nuclear physics to include the effect of three-nucleon forces through the normal ordered two-body approximation. A key improvement needed in order to achieve this progress was an appropriate storage scheme for the nuclear matrix elements in momentum space. We have applied this extended framework to the calculations of properties of pure neutron matter such as the equation of state, the one-body momentum distribution and the static structure factor. The accurate calculation of the latter two quantities was made possible by representing the many-body state in a Slater determinant basis formulated directly in momentum space. While critical for both the energy per particle and, to some extent, for the one-body momentum distribution we found that the three-nucleon interaction has only a modest effect on the static structure factor at nuclear saturation density. This is a first step towards fully \emph{ab initio} simulations of the properties of bulk nuclear matter starting from chiral nuclear interactions and lays the groundwork for understanding the systematic effects introduced by the necessary regulators. In a follow-up work we plan to directly compare both local and non-local low-momentum regulators and study how the breaking of Fierz symmetry impacts the extraction of properties of neutron matter at densities higher than nuclear saturation. This is a unique feature which sets aside CIMC with respect to more conventional Quantum Monte Carlo methods formulated in coordinate space which are instead limited to local regulators and for which the inclusion of interactions beyond next-to-next-to-leading-order requires more sophisticated approaches.
\label{sec:conclusions}

\begin{acknowledgments}
The authors are grateful to A.~Rios for help benchmarking their code and to A.~Schwenk and V.~Somà for comments on the manuscript. 
This work is supported in part by the UK Science and Technology Facilities Council (STFC) through grants No.~ST/L005816/1 and No.~ST/V001108/1,
the  Deutsche  Forschungsgemeinschaft  (DFG,  German Research Foundation) -- Project-ID 279384907 -- SFB 1245, by the BMBF Contract No.~05P18RDFN1 and by the U.S. Department of Energy (DOE), Office of Science, Office of Nuclear Physics, Inqubator for Quantum Simulation (IQuS) under Award Number DOE (NP) Award DE-SC0020970.
This research used resources of the National Energy Research
Scientific Computing Center, a DOE Office of Science User Facility
supported by the Office of Science of the U.S. Department of Energy
under Contract No.~DE-AC02-05CH11231 using NERSC award
NP-ERCAP0020946 and resources at the DiRAC DiAL system at the University of Leicester, UK, (funded  by  the  UK  BEIS via  STFC  Capital  Grants No.~ST/K000373/1 and No.~ST/R002363/1 and STFC DiRAC  Operations  Grant  No.~ST/R001014/1).

\end{acknowledgments}

\appendix

\section{Matrix element storage}
\label{sec:ME_storage}

\begin{figure}[t]
  \includegraphics[width=\linewidth]{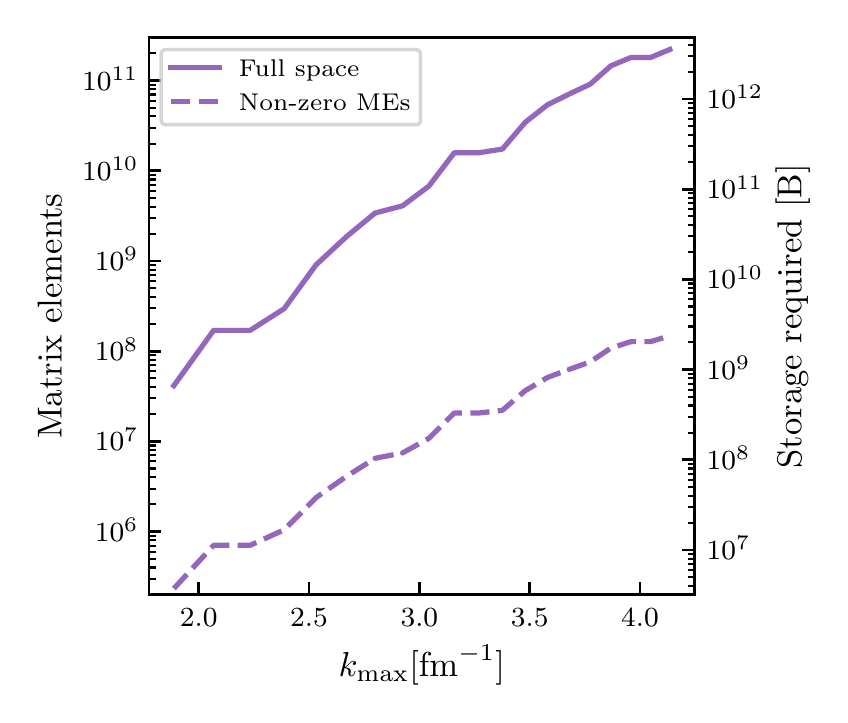}%
\caption{\label{fig:me_storage} Number of matrix elements depending on whether the full accessible configuration space or only the non-zero normal-ordered matrix elements for \opt with 66 neutrons at $\rho_0=0.16$ fm$^{-3}$ are considered, as well as the corresponding storage space required in double precision.}
\end{figure}

One of the main perks of CIMC is that working directly in configuration space, the on-the-fly computation of non-local $\chi$-EFT matrix elements is easy and fast, and this does not require the pre-processing of the matrix elements needed by other many-body methods working in a harmonic oscillator basis. While this on-the-fly approach was previously used in CIMC, its computational cost rises tremendously when incorporating normal-ordered 3NF, as the matrix elements now include a summation over the third particle, scaling with the number of nucleons included in the simulation. This makes the calculations intractable, and thus the matrix elements need to be computed once and stored in memory.

But as the number of normal-ordered two-body matrix elements grows as $\mathcal{O}(N^4_{s.p.})$, with $N_{s.p.}$ the size of the single-particle model space, realistic calculations using large model spaces come with storage requirements that exceed the capacity of modern clusters, as displayed on Fig.~\ref{fig:me_storage}. The obvious choice is then to store only the non-zero matrix elements $N_{n.z.}$. This tremendously reduces the matrix elements storage requirement, but a map from the single-particle configurations indices to the non-zero matrix elements ones still needs to be stored. As this map scales as $\mathcal{O}(N^4_{s.p.})$ as well, one uses a hash table scaling as $\mathcal{O}(N_{n.z.})$ instead, 
with matrix elements being shared over different MPI processes but keeping a full copy of all non-zero elements on each separate computing node.
This ensures that the memory requirement remains tractable and all matrix elements can be stored for the calculations.

\section{Computation of the response function}
\label{sec:Sk_formalism}

We now provide technical details of our implementation of the static structure factor $S(q)$ within the CIMC approach. We perform a random walk with importance sampling using a state $\rvert\Phi_G\rangle$ as guiding function. The projected state at time $\tau$ is
\begin{equation}
\ket{\Psi_G(\tau)} = \sum_{l=1}^{N_\tau} w_G(l,\tau)\ket{D_l}\;,
\end{equation}
with $w_G(l,\tau)\in\mathbb{R}^+$ a set of weights, $N_\tau$ their number at time $\tau$, and $\ket{D_l}$ a $N$-particle Slater determinant. We will consider the mixed estimator
\begin{equation}
S_{mixed}({\bf q},\tau) = \frac{\langle \Phi_G\lvert \hat{S}({\bf q})\rvert \Psi(\tau)\rangle}{\langle \Phi_G\vert \Psi(\tau)\rangle}\;,
\end{equation}
where ${\bf q}$ is a three-dimensional vector and
\begin{equation}
\begin{split}
\hat{S}({\bf q})
&=\frac{1}{N}\sum_{\sigma,\sigma'}\sum_{{\bf k},{\bf k}'} a^\dagger_{{\bf k}-{\bf q},\sigma}a_{{\bf k},\sigma}a^\dagger_{{\bf k}'+{\bf q},\sigma'}a_{{\bf k}',\sigma'}\;.
\end{split}
\end{equation}
The state $\ket{\Psi(\tau)}$ is obtained by removing the guiding function from $\ket{\Psi_G(\tau)}$, in components
\begin{equation}
\langle D_m\vert\Psi(\tau)\rangle = \frac{\langle D_m\vert\Psi_G(\tau)\rangle}{\langle\Phi_G\vert D_m\rangle}
\end{equation}
for a determinant $\ket{D_m}$. The mixed expectation value can then be written more explicitly as follows
\begin{equation}
S_{mixed}({\bf q},\tau) = \frac{\sum_{l=1}^{N_\tau}w_G(l,\tau)\frac{\langle \Phi_G\lvert \hat{S}({\bf q})\rvert D_l\rangle}{\langle\Phi_G\vert D_l\rangle}}{\sum_{l=1}^{N_\tau}w_G(l,\tau)}\;,
\end{equation}
from which we can define the local expectation value
\begin{equation}
S_l({\bf q}) = \frac{\langle \Phi_G\lvert \hat{S}({\bf q})\rvert D_l\rangle}{\langle\Phi_G\vert D_l\rangle}\;.
\end{equation}
It is useful to introduce a complete set of determinants and split $S_l({\bf q})$ into a diagonal and an off-diagonal contribution
\begin{equation}
\begin{split}
S_l({\bf q}) &= \langle D_l\rvert \hat{S}({\bf q})\rvert D_l\rangle + \sum_{m\neq l}\frac{\langle \Phi_G\lvert D_m\rangle\langle D_m\rvert \hat{S}({\bf q})\rvert D_l\rangle}{\langle\Phi_G\vert D_l\rangle}\\
&= S^D_l({\bf q}) + S^O_l({\bf q})\;.
\end{split}
\end{equation}

The off-diagonal contribution can be evaluated by restricting the sum of states to the two-particle-two-hole excitations using $\ket{D_l}$ as reference
\begin{equation}
\begin{split}
S^O_l({\bf q}) &= \sum_{a<b}\sum_{i<j} \frac{\langle \Phi_G\lvert ij;ab\rangle}{\langle \Phi_G\lvert D_l\rangle}\langle ij;ab\rvert \hat{S}({\bf q})\rvert D_l\rangle\\
\end{split}
\end{equation}
or, by introducing tensor notation
\begin{equation}
\label{eq:sofq_ordered}
\hat{S}({\bf q}) = \sum_{\alpha,\beta,\gamma,\delta} S_{\alpha \beta,\gamma \delta} \; c^\dagger_\alpha c^\dagger_\beta c_\gamma c_\delta\;,
\end{equation}
equivalently expressed as follows
\begin{equation}
S^O_l({\bf q}) = \sum_{a<b}\sum_{i<j} \frac{\langle \Phi_G\lvert ij;ab\rangle}{\langle \Phi_G\lvert D_l\rangle}\left(S_{a b,i j}-S_{a b,j i}\right)\;.
\end{equation}
In order to bring $\hat{S}({\bf q})$ to the ordered form in Eq.~\eqref{eq:sofq_ordered} we use
\begin{equation}
\begin{split}
\hat{S}({\bf q}) &= \frac{1}{N}\sum_{\sigma,\sigma'}\sum_{{\bf k},{\bf k}'} c^\dagger_{{\bf k}-{\bf q},\sigma}c_{{\bf k},\sigma}c^\dagger_{{\bf k}'+{\bf q},\sigma'}c_{{\bf k}',\sigma'}\\
&=\frac{1}{N} \sum_{\sigma'}\sum_{{\bf k}'} c^\dagger_{{\bf k}',\sigma'}c_{{\bf k}',\sigma'} \\
&-\frac{1}{N}\sum_{\sigma,\sigma'}\sum_{{\bf k},{\bf k}'} c^\dagger_{{\bf k}-{\bf q},\sigma}c^\dagger_{{\bf k}'+{\bf q},\sigma'}c_{{\bf k},\sigma}c_{{\bf k}',\sigma'}\\
&=\frac{\hat{N}}{N}-\frac{1}{N}\sum_{\sigma,\sigma'}\sum_{{\bf k},{\bf k}'} c^\dagger_{{\bf k}-{\bf q},\sigma}c^\dagger_{{\bf k}'+{\bf q},\sigma'}c_{{\bf k},\sigma}c_{{\bf k}',\sigma'}\;,
\end{split}
\end{equation}
where we used the definition of the (total) number operator. The first term is diagonal and does not contribute to $S^O_l({\bf q})$ while the second one can be written as in Eq.~\eqref{eq:sofq_ordered} by choosing
\begin{equation}
\begin{split}
S_{\alpha \beta,\gamma \delta} &= -\frac{1}{N}\delta_{\sigma_\alpha,\sigma_\gamma}\delta_{\sigma_\beta,\sigma_\delta} \\
&\phantom{=} \times \delta^{(3)}({\bf k}_\alpha-{\bf k}_\gamma+{\bf q})\delta^{(3)}({\bf k}_\beta-{\bf k}_\delta-{\bf q})\;.
\end{split}
\end{equation}
For the diagonal contribution we find instead that
\begin{equation}
\label{eq:sofq_diag}
S^D_l({\bf q}) = N \delta^{(3)}({\bf q}) + \frac{1}{N}\sum_\sigma\sum_{\bf k} \left(1-n^l_{{\bf k}-{\bf q},\sigma}\right)n^l_{{\bf k},\sigma}\;,
\end{equation}
with the occupation numbers
\begin{equation}
n^l_{{\bf k},\sigma} = \langle D_l\lvert a^\dagger_{{\bf k},\sigma}a_{{\bf k},\sigma} \rvert D_l\rangle.
\end{equation}
The expression for the diagonal part in Eq.~\eqref{eq:sofq_diag} shows already that, for $|{\bf q}|>2k_F$, the structure factor is exactly equal to 1 for the Hartree-Fock state while $m$-particle $m$-hole states can acquire a correction of order $O(m/N)$. In order to account for shell effects generated by using a finite simulation box with $N$ particles, we employ the following simple approach. We assume the finite-size distortion of the static structure factor can be modelled by a momentum-dependent multiplicative factor
\begin{equation}
S^{(N)}(q) = S(q)\left(1+\epsilon^{(N)}(q)\right) \;,
\label{eq:eq_corr}
\end{equation}
and that this perturbation is mostly independent of interactions. We can then estimate $\epsilon^{(N)}$ by taking the ratio between the free gas response function $S_0(q)$ with and without finite-size effects
\begin{equation}
\left(1+\epsilon^{(N)}(q)\right) = S^{(N)}_0(q)/S_0(q) \;.
\label{eq:eq_mf}
\end{equation}
\begin{figure}[t]
  \includegraphics[width=\linewidth]{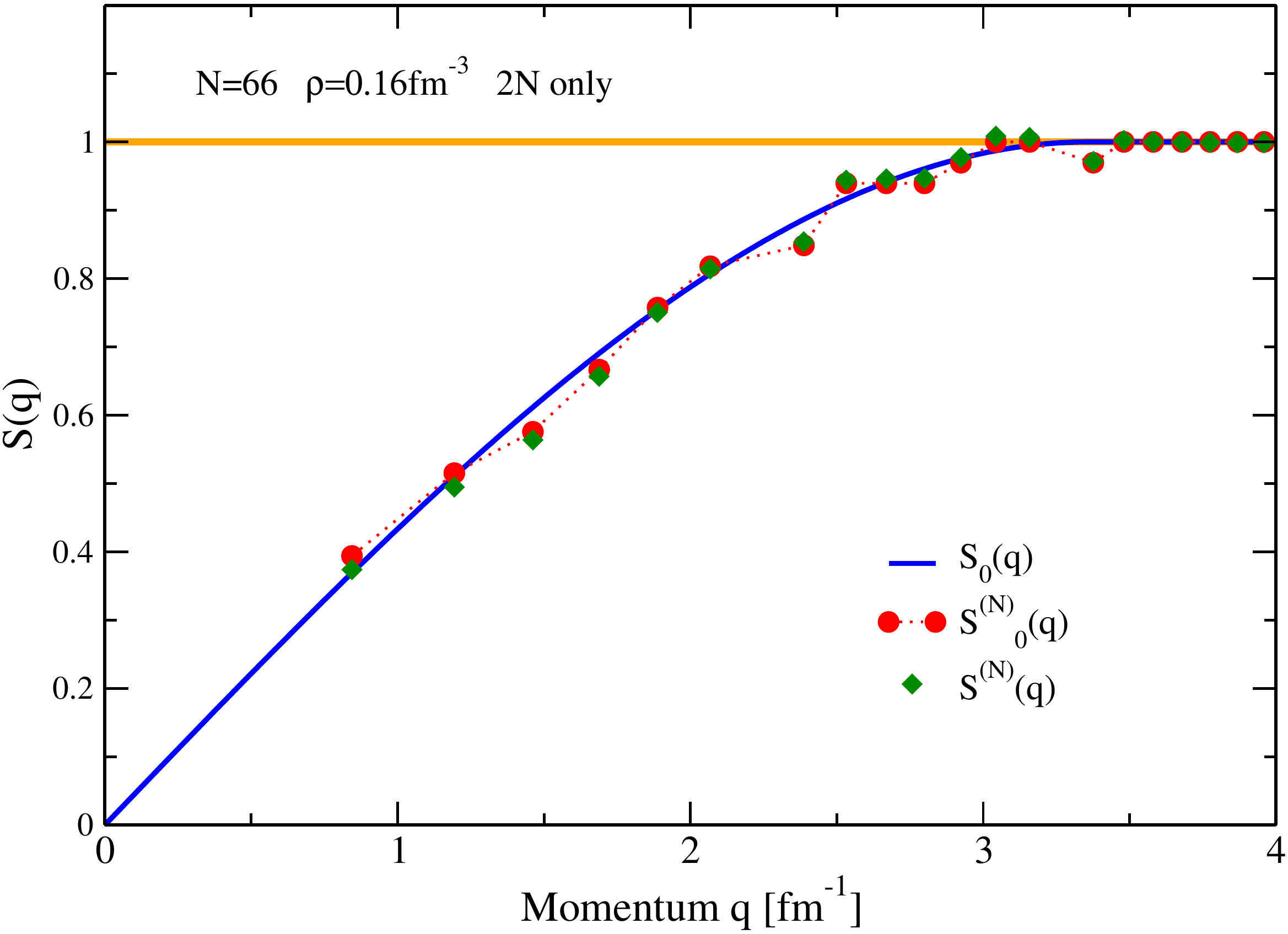}
  \caption{\label{fig:sofq_fs} Free particle structure factors $S_0(q)$ and $S^{(N)}_0(q)$ for $N=66$ neutrons at saturation density. Also shown is the structure factor $S^{(N)}(q)$ for the \opt Hamiltonian with NN forces only.}
\end{figure}
The corrected static response reported in Fig.~\ref{fig:sofq} of the main text is then obtained as
\begin{equation}
S(q) = \frac{S_0(q)}{S^{(N)}_0(q)}S^{(N)}(q) \;.
\end{equation}
This correction is important for a system of only 66 nucleons as shell effects are still sizeable and stronger than the contribution of interactions. For reference we show in Fig.~\ref{fig:sofq_fs} the free particle response functions $S^{(N)}_0(q)$ (red circles), $S_0(q)$ (full blue line) and $S^{(N)}(q)$ (green diamonds) for $\rho = 0.16\ \mathrm{fm}^{-3}$ with NN forces only.  
The very close trends for $S^{(N)}_0(q)$ and $S^{(N)}(q)$ confirm our assumption that finite size effects are largely independent of correlations and therefore the same values of $\epsilon^{(N)}(q)$
can enter both Eqs.~\eqref{eq:eq_corr} and~\eqref{eq:eq_mf}.

\bibliography{References}

\end{document}